\documentclass[prd,aps,showpacs,tightenlines,superscriptaddress,amsmath,amssymb,amsfonts]{revtex4-2}
\usepackage{amssymb,amsmath,amsthm,amsbsy,epsfig,color,graphicx,times}
\usepackage[ansinew]{inputenc}
\usepackage[english]{babel}
\usepackage{color}
\usepackage{braket}
\usepackage{tabularx}
\usepackage{array}
\usepackage{slashed}
\begin{document}

\title{The $X_{17}$ Anomaly: Experimental Evidence and Theoretical Interpretations}

\author{R. Serao}
\email{rserao@unisa.it}
\affiliation{Dipartimento di Fisica ``E.R. Caianiello'' Universit\`{a} di Salerno, and INFN --- Gruppo Collegato di Salerno, Via Giovanni Paolo II, 132, 84084 Fisciano (SA), Italy}

\author{A. Quaranta}
\email{aniello.quaranta@unicam.it}
\affiliation{School of Science and Technology, University of Camerino,
Via Madonna delle Carceri, 62032 Camerino, Italy}

\author{A. Capolupo}
\email{capolupo@sa.infn.it}
\affiliation{Dipartimento di Fisica ``E.R. Caianiello'' Universit\`{a} di Salerno, and INFN --- Gruppo Collegato di Salerno, Via Giovanni Paolo II, 132, 84084 Fisciano (SA), Italy}

\begin{abstract}
This review summarizes the experimental evidence for the hypothetical $X_{17}$  particle \cite{X2,X3}, examines the theoretical frameworks in which it can be accommodated \cite{Feng1,Feng2}, and discusses its potential implications for the Standard Model and couplings to known particles \cite{CAPOLUPO}. Future experimental prospects are also highlighted.
\end{abstract}

\maketitle

\section{Introduction}

The Standard Model (SM) of particle physics represents one of the most successful theoretical frameworks in modern physics, providing an accurate description of fundamental particles and their interactions through electromagnetic, weak, and strong forces. Its predictions have been extensively validated, culminating in the discovery of the Higgs boson at the Large Hadron Collider (LHC) \cite{higgs2012}. 

Despite its success, the SM is widely regarded as an effective theory, expected to be incomplete. Several theoretical and experimental observations point toward the existence of physics beyond the Standard Model (BSM). Among the most notable issues are the absence of a viable dark matter candidate \cite{darkmatter1,darkmatter2,darkmatter3,darkmatter4,darkmatter5,darkmatter6,darkmatter7,darkmatter8,darkmatter9}, the origin of neutrino masses and flavor mixing \cite{seesaw, mixing1,mixing2,mixing2bis,mixing2bbis,mixing3,mixing4,mixing5,mixing6,agg1,agg2,agg3,agg4,agg5},, and the inability to account for the observed baryon asymmetry of the Universe \cite{asymmetry1,asymmetry2,asymmetry3}. In addition, the SM does not address the structure of fermion masses and mixings, nor does it incorporate gravity in a consistent quantum framework.

While high-energy collider experiments probe BSM physics at the TeV scale, low-energy precision measurements provide a complementary approach, particularly sensitive to light, weakly coupled particles. In this context, recent nuclear physics experiments have reported intriguing anomalies that may hint at the existence of new light degrees of freedom.

In particular, the ATOMKI collaboration has observed an unexpected excess in the angular correlations of electron-positron ($e^+e^-$) pairs produced in nuclear transitions of excited states of $^8$Be \cite{X2,X3} and later in $^4$He \cite{X4}. These results can be interpreted as the production of a new neutral boson with a mass of approximately $17$ MeV, commonly referred to as $X_{17}$  \cite{X5,X6}. The kinematic features of the observed signal are not easily explained within standard nuclear physics or quantum electrodynamics, making this anomaly a potential indication of new physics \cite{agg6,agg7,agg8}.

However, the existence of the $X_{17}$ particle remains highly debated. Independent experimental searches have not yet provided conclusive confirmation, and several constraints from particle physics experiments, including beam dump experiments and precision measurements, significantly restrict the allowed parameter space. As a result, any viable interpretation of the $X_{17}$ anomaly must reconcile the observed nuclear signals with stringent bounds from other sectors.

On the theoretical side, various models have been proposed to explain the anomaly. These include protophobic vector bosons \cite{Feng1,Feng2}, light scalar or pseudoscalar particles, and scenarios involving portals to a dark sector. Each of these frameworks faces different challenges in satisfying experimental constraints while reproducing the observed signal.

In this work, we provide a focused review of the experimental and theoretical status of the $X_{17}$  anomaly, with particular emphasis on its phenomenological implications. In addition to summarizing the current state of the art, we present a unified analysis of the constraints on the couplings of the $X_{17}$ boson arising from precision observables. In particular, we consider contributions to the anomalous magnetic moments of leptons, Lamb shift measurements in muonic atoms, and electroweak precision observables. This combined approach allows us to identify the regions of parameter space where the $X_{17}$  hypothesis remains viable and to highlight the interplay between different experimental probes.

The paper is structured as follows. In Sec.~II we review the experimental evidence and current constraints. In Sec.~III we discuss the main theoretical frameworks proposed in the literature. In Sec.~IV we present a detailed phenomenological analysis of the $X_{17}$  couplings and their implications for precision observables. Finally, we summarize our conclusions in Sec.~V.

\section{The ATOMKI experiment and other constraints}

The ATOMKI experiment investigates internal pair creation (IPC) processes in nuclear transitions of light nuclei, searching for deviations from the well-established predictions of quantum electrodynamics (QED). In IPC, an excited nucleus emits a virtual photon that converts into an electron-positron ($e^+e^-$) pair. The angular correlations and invariant mass distributions of these pairs are precisely predicted for given nuclear transitions, making IPC a sensitive probe of possible new light particles \cite{X2,X3,X4,X5,X6}.

The key reaction studied is $^7\mathrm{Li}(p,\gamma)^8\mathrm{Be}$, where a proton beam with energy around $1.03$ MeV selectively populates the $18.15$ MeV $1^+$ isoscalar excited state of $^8$Be. This state decays to the ground state via gamma emission or IPC \cite{X2,X3}. Due to its isoscalar nature, this transition suppresses standard electromagnetic contributions, enhancing sensitivity to potential new bosons.

In measurements of the angular correlation of emitted $e^+e^-$ pairs, the ATOMKI collaboration reported a statistically significant excess at large opening angles (around $140^\circ$) compared to QED expectations. This feature appears as a bump in the reconstructed invariant mass distribution, corresponding to a mass of approximately $m_X \simeq 16.7$ MeV/$c^2$, with a quoted significance above $6\sigma$. A similar anomaly was later reported in transitions of $^4$He \cite{X4}, strengthening the interest in a possible new particle interpretation.

The observed excess can be interpreted as the production of a new neutral boson, denoted as $X_{17}$ , emitted in a two-body decay process $^8\mathrm{Be}^* \rightarrow ^8\mathrm{Be} + X$, followed by $X \rightarrow e^+e^-$. This mechanism leads to kinematic distributions distinct from standard IPC processes.

Despite these intriguing observations, independent experimental verification remains inconclusive. The MEG II collaboration performed a dedicated search for similar signals in electron-positron angular correlations but did not observe any statistically significant deviation from the expected background \cite{MEGII}. While their results do not fully exclude the parameter region suggested by ATOMKI, they do not provide supporting evidence for the anomaly.

On the other hand, a measurement performed by the VNU University of Science reported an excess compatible with the $X_{17}$  interpretation in the reaction $^7\mathrm{Li}(p,e^+e^-)^8\mathrm{Be}$, with a reconstructed mass $m_X = 16.66 \pm 0.47\,(\mathrm{stat}) \pm 0.35\,(\mathrm{sys})$ MeV and a significance exceeding $4\sigma$ \cite{VNU}. However, this result requires independent confirmation.

Further constraints arise from beam dump and fixed-target experiments. In particular, the NA64 collaboration at CERN searched for the production of a light boson in the bremsstrahlung process $e^- Z \rightarrow e^- Z X$, followed by $X \rightarrow e^+e^-$ \cite{NA64}. No signal was observed, leading to stringent bounds on the coupling of $X_{17}$  to electrons, significantly restricting the allowed parameter space.

Overall, the current experimental situation remains inconclusive. While the ATOMKI results and some independent measurements suggest the possible existence of a new light boson, other dedicated searches have not confirmed the anomaly. This tension highlights the need for further high-precision experiments to clarify the nature of the observed excess and to determine whether it can be attributed to new physics or to unresolved nuclear or experimental effects.

\section{Theoretical framework}

The ATOMKI anomaly has triggered significant theoretical activity aimed at constructing consistent extensions of the Standard Model capable of accommodating a light boson with mass around $17$ MeV. Any viable framework must simultaneously reproduce the observed nuclear decay signatures while satisfying the stringent constraints from particle physics experiments, astrophysics, and cosmology. This requirement severely restricts the class of admissible models.

Among the proposed explanations, the most extensively studied scenario involves a new light vector boson with suppressed couplings to protons, often referred to as a protophobic vector boson \cite{Feng1,Feng2}. In this framework, the coupling structure is arranged such that interactions with neutrons dominate over those with protons, allowing the model to evade strong bounds from electron scattering and other precision measurements. The required effective couplings are typically of order $10^{-4}$--$10^{-3}$, which are sufficient to account for the observed signal while remaining compatible with existing constraints. Furthermore, angular momentum conservation in the nuclear transition $^8\mathrm{Be}^* \rightarrow ^8\mathrm{Be} + e^+e^-$ disfavors a scalar interpretation, making a vector mediator a more consistent candidate.

Alternative explanations based on scalar or pseudoscalar particles, such as axion-like states or pseudo-Nambu-Goldstone bosons \cite{Ell}, have also been considered. However, these scenarios are strongly constrained by bounds from meson decays, beam dump experiments, and astrophysical observations, making it challenging to construct models that simultaneously allow for efficient production in nuclear transitions and satisfy all existing limits.

More general approaches involve the introduction of portals to a dark sector, where a light mediator connects Standard Model particles to additional hidden degrees of freedom \cite{ALves}. These include vector portals, typically realized through kinetic mixing with a new $U(1)'$ gauge symmetry, as well as scalar portals involving couplings to the Higgs sector. While such frameworks offer a broader phenomenological landscape, they are subject to significant experimental constraints and often require nontrivial parameter tuning to remain viable.

Given these considerations, in the following we focus on a generic phenomenological description of a light vector mediator with effective couplings to Standard Model fermions. The corresponding Lagrangian can be written as:
\begin{equation}
    \mathcal{L}=-\frac{1}{4}X_{\mu\nu}X^{\mu\nu}+\frac{1}{2}m_X^2 X_\mu X^\mu - X_\mu J^\mu,
\end{equation}
where $X_{\mu\nu}$ is the field strength tensor of the new gauge boson and the current $J^\mu$ encodes its couplings to fermions. At the fundamental level, one can write
\begin{equation}
    J^\mu = \sum_f e\, \epsilon_f \, \bar{f}\gamma^\mu f,
\end{equation}
with $f = e,\mu,\tau$, while at the nucleon level
\begin{equation}
    J^\mu_N = e\,\epsilon_p\,\bar{p}\gamma^\mu p + e\,\epsilon_n\,\bar{n}\gamma^\mu n.
\end{equation}
The effective couplings to protons and neutrons are related to the underlying quark couplings through $\epsilon_p = 2\epsilon_u + \epsilon_d$ and $\epsilon_n = \epsilon_u + 2\epsilon_d$. Constraints on these parameters can be found in \cite{Feng1,Feng2}.

In the next section, we will use this effective framework to analyze the phenomenological implications of the $X_{17}$  boson, focusing on its couplings to leptons and nucleons and their impact on precision observables.

\section{Phenomenology of the $X_{17}$ boson}

The existence of a light vector boson with mass $M_X \simeq 17$ MeV and weak couplings to Standard Model fermions can lead to observable effects in a variety of precision measurements. In this section, we analyze the phenomenological implications of the $X_{17}$ boson, focusing on its contributions to leptonic anomalous magnetic moments, atomic physics observables, and electroweak precision tests. These complementary probes allow us to constrain the coupling structure of the model and assess the viability of the $X_{17}$ hypothesis.

\subsection{Lepton anomalous magnetic moments}

One of the most sensitive probes of new light particles is provided by the anomalous magnetic moments of charged leptons. In particular, the long-standing discrepancy between the experimental measurement and the Standard Model prediction of the muon anomalous magnetic moment, $a_\mu = (g-2)_\mu/2$, makes it a natural observable to test the presence of a light vector mediator.

In the presence of the $X_{17}$ boson, additional contributions arise at one-loop level through vertex corrections. The interaction between the $X_{17}$ and leptons is described by the effective current
\begin{equation}
J^\mu = \sum_l e\,\epsilon_l\,\bar{l}\gamma^\mu l,
\end{equation}
which leads to a vertex factor $-i e \epsilon_l \gamma^\mu$.

The scattering amplitude of a lepton in an external electromagnetic field can be written as
\begin{equation}
i \mathcal{M} = i e \, \bar{u}(p') \Gamma^\mu(p',p) u(p)\, A_\mu(q),
\end{equation}
where the vertex function admits the general decomposition
\begin{equation}
\Gamma^\mu(p',p) = F_1(q^2)\gamma^\mu + F_2(q^2)\frac{i\sigma^{\mu\nu}q_\nu}{2m_l}.
\end{equation}
The anomalous magnetic moment is determined by the Pauli form factor as $a_l = F_2(0)$.

The leading contribution of the $X_{17}$ boson arises from the one-loop diagram involving the exchange of the vector mediator. The corresponding correction to the vertex function can be written as
\begin{equation}
\delta \Gamma^\mu = \int \frac{d^4 k}{(2\pi)^4} \bar{u}(p') (-i e \epsilon_l \gamma^\alpha)
\frac{i(\slashed{p}'+\slashed{k}+m_l)}{(p'+k)^2 - m_l^2}
(e\gamma^\mu)
\frac{i(\slashed{p}+\slashed{k}+m_l)}{(p+k)^2 - m_l^2}
(-i e \epsilon_l \gamma^\beta)
\frac{-i}{k^2 - M_X^2}
\left(g_{\alpha\beta} - \frac{k_\alpha k_\beta}{M_X^2}\right)
u(p).
\end{equation}

The term proportional to $k_\alpha k_\beta$ does not contribute to the anomalous magnetic moment after integration \cite{anch}, and the relevant contribution can be extracted from the part proportional to $g_{\alpha\beta}$. After standard Feynman parametrization and loop integration, the contribution of the $X_{17}$ boson to the anomalous magnetic moment of a lepton $l$ can be expressed as
\begin{equation}
a_l^X = \frac{\alpha}{2\pi} \epsilon_l^2 \int_0^1 dx \, \frac{2x^2(1-x)m_l^2}{m_l^2 x^2 + M_X^2 (1-x)}.
\end{equation}

Introducing the dimensionless parameter $\lambda_l = m_l^2 / M_X^2$, this expression can be rewritten as
\begin{equation}
a_l^X = \frac{\alpha}{2\pi} \epsilon_l^2 \, f(\lambda_l),
\end{equation}
where
\begin{equation}
f(\lambda_l) = \lambda_l \int_0^1 dx \, \frac{2x^2(1-x)}{\lambda_l x^2 - x + 1}.
\end{equation}

This result highlights the explicit dependence of the correction on the ratio between the lepton mass and the mediator mass. As a consequence, the effect is strongly suppressed for the electron, while it becomes relevant for the muon and potentially sizable for the tau lepton.

Assuming that the observed discrepancy in the muon anomalous magnetic moment is entirely due to the $X_{17}$ contribution, one can impose the constraint
\begin{equation}
a_\mu^X \lesssim \Delta a_\mu = a_{\mu,\mathrm{EXP}} - a_{\mu,\mathrm{SM}} \simeq 2.51 \times 10^{-9},
\end{equation}
which leads to an upper bound on the coupling
\begin{equation}
|\epsilon_\mu| \lesssim 2.1 \times 10^{-4}.
\end{equation}

Similarly, the much stronger agreement between theory and experiment for the electron anomalous magnetic moment implies a significantly tighter constraint on $\epsilon_e$, although the contribution is suppressed by the small electron mass.

These results indicate that any viable realization of the $X_{17}$ boson must feature non-universal couplings to leptons, as flavor-independent scenarios would lead to tensions with the stringent bounds from the electron anomalous magnetic moment.

\subsection{Lamb shift in muonic atoms}

Precision spectroscopy of muonic atoms provides an additional and highly sensitive probe of new light mediators. In particular, the measured Lamb shift in muonic hydrogen and deuterium exhibits deviations from the Standard Model expectations, which can be used to constrain the couplings of the $X_{17}$ boson.

The exchange of the $X_{17}$ boson between a muon and a nucleon induces an additional Yukawa-type potential of the form
\begin{equation}
V_X(r) = \epsilon_\mu \epsilon_N \, \alpha \, \frac{e^{-M_X r}}{r},
\end{equation}
where $\epsilon_\mu$ and $\epsilon_N$ denote the effective couplings of the $X_{17}$ to the muon and to the nucleon ($N = p,n$), respectively.

The corresponding correction to the atomic energy levels can be computed in first-order perturbation theory. Focusing on the Lamb shift, i.e. the energy difference between the $2S_{1/2}$ and $2P_{3/2}$ states, one obtains
\begin{equation}
\delta E_X = \int_0^\infty dr\, r^2 \, V_X(r) \left( |R_{20}(r)|^2 - |R_{21}(r)|^2 \right),
\end{equation}
where $R_{nl}(r)$ are the radial wave functions of the hydrogen-like system.

For muonic hydrogen, the above expression can be evaluated analytically, leading to
\begin{equation}
\delta E_X^H = \frac{\alpha}{2 a_H^3} \, \epsilon_\mu \epsilon_p \, \frac{f(a_H M_X)}{M_X^2},
\end{equation}
where $a_H = (\alpha m_{\mu p})^{-1}$ is the Bohr radius of muonic hydrogen, $m_{\mu p}$ is the reduced mass, and the dimensionless function
\begin{equation}
f(x) = \frac{x^4}{(1+x)^4}
\end{equation}
encodes the dependence on the mediator mass.

\begin{figure}[t]
\centering
\includegraphics[width=0.6\columnwidth]{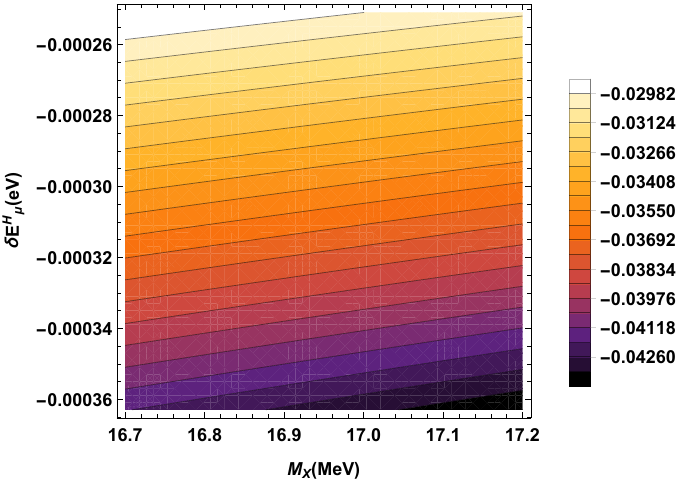}
\caption{
Contour plot of the lower bound for the coupling $\epsilon_p$ as a function of $M_X$ and $\delta E_\mu^H$. We fix $\epsilon_\mu \simeq 2.154 \times 10^{-4}$.
}
\label{fig:epsilon_p}
\end{figure}

Experimental measurements of the Lamb shift in muonic hydrogen report a deviation in the range
\begin{equation}
\delta E_\mu^H \in (-0.363, -0.251)\ \mathrm{meV},
\end{equation}
which can be used to constrain the product of couplings $\epsilon_\mu \epsilon_p$.
The corresponding constraints on the proton coupling are shown in Fig.~\ref{fig:epsilon_p}.

A crucial feature of this result is that the sign of the observed shift requires opposite signs for the muon and proton couplings, i.e.
\begin{equation}
\epsilon_\mu \, \epsilon_p < 0.
\end{equation}
This condition represents a nontrivial constraint on the structure of the underlying theory.

An analogous analysis can be performed for muonic deuterium, leading to constraints on the neutron coupling $\epsilon_n$. In this case, the measured deviation is
\begin{equation}
\delta E_\mu^D \in (-0.475, -0.337)\ \mathrm{meV},
\end{equation}
and the resulting bounds depend on both $\epsilon_p$ and $\epsilon_n$, reflecting the composite nature of the deuteron.

\begin{figure}[t]
\centering
\includegraphics[width=0.6\columnwidth]{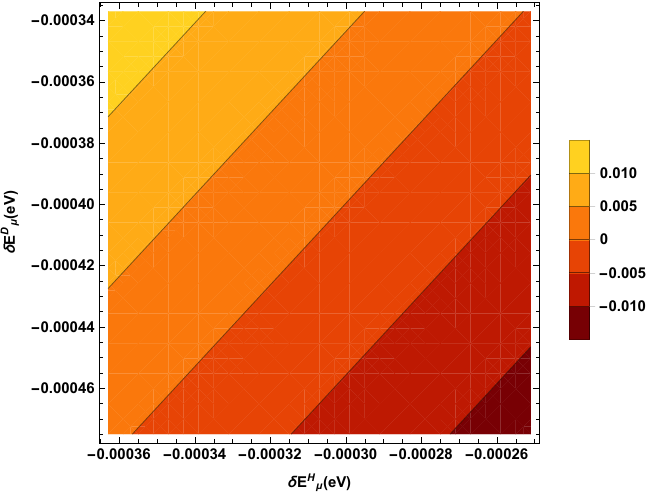}
\caption{
Constraints on the neutron coupling $\epsilon_n$ obtained from muonic hydrogen and deuterium data, assuming $|\epsilon_\mu| \simeq 2.154 \times 10^{-4}$ and $M_X = 17$ MeV.
}
\label{fig:epsilon_n}
\end{figure}

The resulting constraints on the neutron coupling are shown in Fig.~\ref{fig:epsilon_n}.

Combining the Lamb shift constraints with those obtained from the anomalous magnetic moment, one obtains complementary information on the coupling structure of the $X_{17}$ boson. In particular, while the $g-2$ observable constrains $\epsilon_\mu$ independently, the Lamb shift probes directly the product $\epsilon_\mu \epsilon_N$, allowing for a determination of the relative sign and magnitude of the couplings.

These results highlight the importance of atomic physics observables in testing light mediator scenarios and provide stringent bounds that must be satisfied by any viable realization of the $X_{17}$ hypothesis.

\subsection{Kinetic mixing and electroweak precision constraints}

In addition to low-energy observables, light vector mediators can also affect electroweak precision measurements through kinetic mixing with the Standard Model gauge bosons. In particular, a $U(1)'$ gauge boson such as the $X_{17}$ generically admits a kinetic mixing term with the hypercharge field, which can be written as
\begin{equation}
\mathcal{L} \supset -\frac{1}{4}\hat{B}_{\mu\nu}\hat{B}^{\mu\nu}
-\frac{1}{4}\hat{X}_{\mu\nu}\hat{X}^{\mu\nu}
+\frac{\xi}{2c_W}\hat{X}_{\mu\nu}\hat{B}^{\mu\nu}
+\frac{1}{2}M_{X,0}^2 \hat{X}_\mu \hat{X}^\mu,
\end{equation}
where $\xi$ is the kinetic mixing parameter and $c_W = \cos\theta_W$.

Even if absent at tree level, such a term is typically generated radiatively, making it a generic feature of these models. After diagonalization and canonical normalization of the gauge fields, the mixing induces shifts in the electroweak gauge boson masses and couplings.

At leading order, these effects can be parametrized in terms of the oblique parameters $S$, $T$, and $U$, which in the presence of kinetic mixing take the form \cite{W03,W2}
\begin{equation}
\begin{aligned}
S &= \frac{4 s_W^2 (c_W^2 - r^2)}{\alpha(M_Z^2)} \left( \frac{\xi}{1 - r^2} \right)^2, \\
T &= -\frac{s_W^2 r^2}{c_W^2 \alpha(M_Z^2)} \left( \frac{\xi}{1 - r^2} \right)^2, \\
U &= \frac{4 s_W^4}{\alpha(M_Z^2)} \left( \frac{\xi}{1 - r^2} \right)^2,
\end{aligned}
\end{equation}
where $r = M_X / M_Z$.

These corrections induce a shift in the $W$ boson mass given by
\begin{equation}
M_W^2 = (M_W^{\mathrm{SM}})^2 \left[1 + \frac{\alpha c_W^2}{c_W^2 - s_W^2}
\left(-\frac{1}{2}S + c_W^2 T + \frac{c_W^2 - s_W^2}{4 s_W^2} U \right)\right].
\end{equation}

For $M_X \ll M_Z$, the dominant effect simplifies and leads to a negative shift in the $W$ mass:
\begin{equation}
\Delta M_W \simeq -\frac{M_W^{\mathrm{SM}} s_W^2}{2 (c_W^2 - s_W^2)} \, \xi^2.
\end{equation}

Using the current experimental uncertainty on the $W$ boson mass, $\Delta M_W \lesssim 10~\mathrm{MeV}$ \cite{W1}, one can derive an upper bound on the kinetic mixing parameter:
\begin{equation}
|\xi| \lesssim 2 \times 10^{-2}.
\end{equation}

This constraint is complementary to those obtained from low-energy observables. While the anomalous magnetic moment and Lamb shift measurements probe directly the couplings of the $X_{17}$ boson to leptons and nucleons, electroweak precision observables constrain its mixing with the Standard Model gauge sector.

Taken together, these results significantly restrict the viable parameter space of the $X_{17}$ scenario, highlighting the interplay between different experimental probes in testing light mediator models.

For clarity, the different observables considered in this analysis and the corresponding combinations of couplings they constrain are summarized in Table~\ref{tab:constraints_summary}.
\begin{table}[t]
\centering
\begin{tabular}{c c c}
\hline
Observable & Coupling combination probed & Main implication \\
\hline
Lepton anomalous magnetic moments & $\epsilon_l^2$ & Upper bounds on $\epsilon_e$ and $\epsilon_\mu$ \\
Muonic hydrogen Lamb shift & $\epsilon_\mu \epsilon_p$ & Requires $\epsilon_\mu\epsilon_p<0$ \\
Muonic deuterium Lamb shift & $\epsilon_\mu(\epsilon_p+\epsilon_n)$ & Constrains the neutron coupling $\epsilon_n$ \\
Electroweak precision observables & $\xi^2$ & Bounds kinetic mixing with the SM gauge sector \\
\hline
\end{tabular}
\caption{
Summary of the main precision observables considered in this work and the corresponding combinations of $X_{17}$ couplings that they constrain.
}
\label{tab:constraints_summary}
\end{table}

\section{Conclusion}

In this work, we have presented a focused review of the experimental and theoretical status of the $X_{17}$  anomaly, with particular emphasis on its phenomenological implications. While the observations reported by the ATOMKI collaboration and subsequent measurements suggest the possible existence of a new light boson with mass around $17$ MeV, the current experimental situation remains inconclusive, with independent searches providing no definitive confirmation.

On the theoretical side, only a restricted class of models, most notably light vector mediators with non-universal couplings, can accommodate the anomaly while remaining compatible with existing constraints. Alternative scenarios, such as scalar or pseudoscalar explanations, are strongly disfavored by laboratory and astrophysical bounds.

Within this framework, we have performed a unified phenomenological analysis of the couplings of the $X_{17}$ boson using multiple precision observables. In particular, we have shown that the anomalous magnetic moments of leptons constrain directly the coupling $\epsilon_\mu$, while Lamb shift measurements in muonic atoms probe the product $\epsilon_\mu \epsilon_N$, providing information on both the magnitude and relative sign of the couplings. In addition, electroweak precision observables impose independent bounds on the kinetic mixing parameter $\xi$.

The combination of these constraints significantly restricts the viable parameter space of the $X_{17}$ scenario. In particular, any consistent realization requires non-universal lepton couplings and specific sign relations between leptonic and hadronic interactions, while remaining compatible with stringent bounds from precision measurements.

Overall, the $X_{17}$  hypothesis remains an intriguing but unconfirmed possibility. Future high-precision experiments, both in nuclear transitions and in complementary probes, will be essential to determine whether the observed anomalies are the first indication of new physics or the result of unresolved experimental or nuclear effects.

\section*{Acknowledgements}
We acknowledge partial financial support from MUR and INFN, A.C. also acknowledges the COST Action CA1511 Cosmology
and Astrophysics Network for Theoretical Advances and Training
Actions (CANTATA).

\bibliography{bibliografia}

@article{higgs2012,
    author = "Aad, Georges and others",
    collaboration = "ATLAS",
    title = "{Observation of a new particle in the search for the Standard Model Higgs boson with the ATLAS detector at the LHC}",
    eprint = "1207.7214",
    archivePrefix = "arXiv",
    primaryClass = "hep-ex",
    reportNumber = "CERN-PH-EP-2012-218",
    doi = "10.1016/j.physletb.2012.08.020",
    journal = "Phys. Lett. B",
    volume = "716",
    pages = "1--29",
    year = "2012"
}

@article{darkmatter1,
    author = "Rubin, Vera C. and Ford, Jr., W. Kent",
    title = "{Rotation of the Andromeda Nebula from a Spectroscopic Survey of Emission Regions}",
    doi = "10.1086/150317",
    journal = "Astrophys. J.",
    volume = "159",
    pages = "379--403",
    year = "1970"
}

@article{darkmatter2,
    author = "Rubin, V. C. and Thonnard, N. and Ford, Jr., W. K.",
    title = "{Rotational properties of 21 SC galaxies with a large range of luminosities and radii, from NGC 4605 /R = 4kpc/ to UGC 2885 /R = 122 kpc/}",
    doi = "10.1086/158003",
    journal = "Astrophys. J.",
    volume = "238",
    pages = "471",
    year = "1980"
}

@article{darkmatter3,
    author = "Trimble, Virginia",
    title = "{Existence and Nature of Dark Matter in the Universe}",
    doi = "10.1146/annurev.aa.25.090187.002233",
    journal = "Ann. Rev. Astron. Astrophys.",
    volume = "25",
    pages = "425--472",
    year = "1987"
}

@article{darkmatter4,
    author = "Corbelli, Edvige and Salucci, Paolo",
    title = "{The Extended Rotation Curve and the Dark Matter Halo of M33}",
    eprint = "astro-ph/9909252",
    archivePrefix = "arXiv",
    doi = "10.1046/j.1365-8711.2000.03075.x",
    journal = "Mon. Not. Roy. Astron. Soc.",
    volume = "311",
    pages = "441--447",
    year = "2000"
}

@article{darkmatter5,
    author = "Su{\'a}rez, Abril and Robles, Victor H. and Matos, Tonatiuh",
    editor = "Moreno Gonz{\'a}lez, Claudia and Madriz Aguilar, Jos{\'e} Edgar and Reyes Barrera, Luz Marina",
    title = "{A Review on the Scalar Field/Bose-Einstein Condensate Dark Matter Model}",
    eprint = "1302.0903",
    archivePrefix = "arXiv",
    primaryClass = "astro-ph.CO",
    doi = "10.1007/978-3-319-02063-1_9",
    journal = "Astrophys. Space Sci. Proc.",
    volume = "38",
    pages = "107--142",
    year = "2014"
}

@article{darkmatter6,
    author = "Clesse, Sebastien and Garc{\'\i}a-Bellido, Juan",
    title = "{Seven Hints for Primordial Black Hole Dark Matter}",
    eprint = "1711.10458",
    archivePrefix = "arXiv",
    primaryClass = "astro-ph.CO",
    reportNumber = "IFT-UAM-CSIC-17-108, CERN-TH-2017-239",
    doi = "10.1016/j.dark.2018.08.004",
    journal = "Phys. Dark Univ.",
    volume = "22",
    pages = "137--146",
    year = "2018"
}

@article{darkmatter7,
    author = "Dawoodbhoy, Taha and Shapiro, Paul R. and Rindler-Daller, Tanja",
    title = "{Core-envelope haloes in scalar field dark matter with repulsive self-interaction: fluid dynamics beyond the de Broglie wavelength}",
    eprint = "2104.07043",
    archivePrefix = "arXiv",
    primaryClass = "astro-ph.CO",
    doi = "10.1093/mnras/stab1859",
    journal = "Mon. Not. Roy. Astron. Soc.",
    volume = "506",
    number = "2",
    pages = "2418--2444",
    year = "2021"
}

@article{darkmatter8,
    author = "Tulin, Sean and Yu, Hai-Bo",
    title = "{Dark Matter Self-interactions and Small Scale Structure}",
    eprint = "1705.02358",
    archivePrefix = "arXiv",
    primaryClass = "hep-ph",
    doi = "10.1016/j.physrep.2017.11.004",
    journal = "Phys. Rept.",
    volume = "730",
    pages = "1--57",
    year = "2018"
}

@article{darkmatter9,
    author = "Capolupo, Antonio and Pisacane, Gabriele and Quaranta, Aniello and Serao, Raoul",
    title = "{Single arm interferometry to probe the scalar field dark matter}",
    eprint = "2505.13574",
    archivePrefix = "arXiv",
    primaryClass = "hep-ph",
    doi = "10.1103/ypm5-jh2j",
    journal = "Phys. Rev. D",
    volume = "113",
    number = "5",
    pages = "055027",
    year = "2026"
}

@article{mixing1,
    author = "Capolupo, Antonio and Quaranta, Aniello",
    title = "{Neutrino capture on tritium as a probe of flavor vacuum condensate and dark matter}",
    eprint = "2205.09640",
    archivePrefix = "arXiv",
    primaryClass = "hep-ph",
    doi = "10.1016/j.physletb.2023.137776",
    journal = "Phys. Lett. B",
    volume = "839",
    pages = "137776",
    year = "2023"
}

@article{mixing2,
    author = "Capolupo, Antonio and Capozziello, Salvatore and Pisacane, Gabriele and Quaranta, Aniello",
    title = "{Missing matter in galaxies as a neutrino mixing effect}",
    eprint = "2411.17319",
    archivePrefix = "arXiv",
    primaryClass = "hep-ph",
    doi = "10.1016/j.dark.2025.101894",
    journal = "Phys. Dark Univ.",
    volume = "48",
    pages = "101894",
    year = "2025"
}

@article{mixing2bis,
    author = "Capolupo, Antonio and Carloni, Sante and Quaranta, Aniello",
    title = "{Quantum flavor vacuum in the expanding universe: A possible candidate for cosmological dark matter?}",
    eprint = "2111.05051",
    archivePrefix = "arXiv",
    primaryClass = "gr-qc",
    doi = "10.1103/PhysRevD.105.105013",
    journal = "Phys. Rev. D",
    volume = "105",
    number = "10",
    pages = "105013",
    year = "2022"
}

@article{mixing2bbis,
    author = "Capolupo, Antonio and Quaranta, Aniello and Setaro, Pia Antonella",
    title = "{Boson mixing and flavor oscillations in curved spacetime}",
    eprint = "2205.09645",
    archivePrefix = "arXiv",
    primaryClass = "hep-th",
    doi = "10.1103/PhysRevD.106.043013",
    journal = "Phys. Rev. D",
    volume = "106",
    number = "4",
    pages = "043013",
    year = "2022"
}

@article{mixing3,
    author = "Capolupo, A. and Monda, S. and Pisacane, G. and Quaranta, A. and Serao, R.",
    title = "{Dark Universe from QFT Mechanisms and Possible Experimental Probes}",
    doi = "10.3390/universe11050142",
    journal = "Universe",
    volume = "11",
    number = "5",
    pages = "142",
    year = "2025"
}

@article{mixing4,
    author = "Capolupo, Antonio and De Maria, Giuseppe and Monda, Simone and Quaranta, Aniello and Serao, Raoul",
    title = "{Quantum Field Theory of Neutrino Mixing in Spacetimes with Torsion}",
    eprint = "2310.09309",
    archivePrefix = "arXiv",
    primaryClass = "hep-ph",
    doi = "10.3390/universe10040170",
    journal = "Universe",
    volume = "10",
    number = "4",
    pages = "170",
    year = "2024"
}

@article{mixing5,
    author = "Capolupo, Antonio and Quaranta, Aniello and Serao, Raoul",
    title = "{Field Mixing in Curved Spacetime and Dark Matter}",
    doi = "10.3390/sym15040807",
    journal = "Symmetry",
    volume = "15",
    number = "4",
    pages = "807",
    year = "2023"
}

@article{mixing6,
    author = "Serao, Raoul and Torre, Gianpaolo and Capolupo, Antonio",
    title = "{Quantum information meets high-energy physics: probing neutrinos and beyond}",
    eprint = "2510.22625",
    archivePrefix = "arXiv",
    primaryClass = "hep-ph",
    doi = "10.1088/1742-5468/ae1df3",
    journal = "J. Stat. Mech.",
    volume = "2025",
    number = "12",
    pages = "124001",
    year = "2025"
}

@article{seesaw,
    author = "King, S.",
    title = "{Neutrinos with mass}",
    journal = "Frontiers",
    volume = "15",
    pages = "19--21",
    year = "2003"
}

@article{asymmetry1,
    author = "Canetti, Laurent and Drewes, Marco and Shaposhnikov, Mikhail",
    title = "{Matter and Antimatter in the Universe}",
    eprint = "1204.4186",
    archivePrefix = "arXiv",
    primaryClass = "hep-ph",
    reportNumber = "TTK-12-04",
    doi = "10.1088/1367-2630/14/9/095012",
    journal = "New J. Phys.",
    volume = "14",
    pages = "095012",
    year = "2012"
}

@article{asymmetry2,
    author = "Boucenna, S. M. and Morisi, S.",
    title = "{Theories relating baryon asymmetry and dark matter: A mini review}",
    eprint = "1310.1904",
    archivePrefix = "arXiv",
    primaryClass = "hep-ph",
    doi = "10.3389/fphy.2013.00033",
    journal = "Front. in Phys.",
    volume = "1",
    pages = "33",
    year = "2014"
}

@article{asymmetry3,
    author = "Borah, Debasish and Jyoti Das, Suruj and Roshan, Rishav",
    title = "{Baryon asymmetry from dark matter decay}",
    eprint = "2305.13367",
    archivePrefix = "arXiv",
    primaryClass = "hep-ph",
    doi = "10.1103/PhysRevD.108.075025",
    journal = "Phys. Rev. D",
    volume = "108",
    number = "7",
    pages = "075025",
    year = "2023"
}

@article{X2,
    author = "Krasznahorkay, A. J. and Krasznahorkay, A. and Csatl{\'o}s, M. and Csige, L. and Tim{\'a}r, J.",
    title = "{A New Particle is Suggested by Anomalies Observed in Internal e$ ^+ $e$ ^- $ Pair Creation}",
    doi = "10.1007/978-3-031-38477-6_5",
    journal = "Springer Proc. Phys.",
    volume = "392",
    pages = "71--80",
    year = "2024"
}

@article{X3,
    author = "Krasznahorkay, Attila J. and Krasznahorkay, Attila and Csatl{\'o}s, Margit and Csige, L{\'o}r{\'a}nt and T{\'\i}m{\'a}r, J{\'a}nos",
    title = "{A New Particle is Being Born in ATOMKI that Could Make a Connection to Dark Matter}",
    doi = "10.1080/10619127.2022.2100157",
    journal = "Nucl. Phys. News",
    volume = "32",
    number = "3",
    pages = "10--15",
    year = "2022"
}

@article{X4,
    author = "Krasznahorkay, A. J. and Csatl{\'o}s, M. and Csige, L. and Guly{\'a}s, J. and Krasznahorkay, A. and Nyak{\'o}, B. M. and Rajta, I. and Tim{\'a}r, J. and Vajda, I. and Sas, N. J.",
    title = "{New anomaly observed in He4 supports the existence of the hypothetical X17 particle}",
    eprint = "2104.10075",
    archivePrefix = "arXiv",
    primaryClass = "nucl-ex",
    doi = "10.1103/PhysRevC.104.044003",
    journal = "Phys. Rev. C",
    volume = "104",
    number = "4",
    pages = "044003",
    year = "2021"
}

@article{X5,
    author = "Krasznahorkay, A. J. and others",
    title = "{Observation of Anomalous Internal Pair Creation in Be8 : A Possible Indication of a Light, Neutral Boson}",
    eprint = "1504.01527",
    archivePrefix = "arXiv",
    primaryClass = "nucl-ex",
    doi = "10.1103/PhysRevLett.116.042501",
    journal = "Phys. Rev. Lett.",
    volume = "116",
    number = "4",
    pages = "042501",
    year = "2016"
}

@article{X6,
    author = "Krasznahorkay, A. J. and others",
    editor = "Gozdz, Marek",
    title = "{Observation of Anomalous Internal Pair Creation in $^8$Be}",
    doi = "10.5506/APhysPolBSupp.8.597",
    journal = "Acta Phys. Polon. Supp.",
    volume = "8",
    number = "3",
    pages = "597",
    year = "2015"
}

@article{MEGII,
    author = "Afanaciev, K. and others",
    collaboration = "MEG II",
    title = "{Search for the X17 particle in $^{7}\textrm{Li}(\textrm{p},\textrm{e}^+ \textrm{e}^{-}) ^{8}\textrm{Be}$ processes with the MEG II detector}",
    eprint = "2411.07994",
    archivePrefix = "arXiv",
    primaryClass = "nucl-ex",
    doi = "10.1140/epjc/s10052-025-14345-0",
    journal = "Eur. Phys. J. C",
    volume = "85",
    number = "7",
    pages = "763",
    year = "2025"
}

@article{VNU,
    author = "Anh, Tran The and others",
    title = "{Checking the $^{8}$Be Anomaly with a Two-Arm Electron Positron Pair Spectrometer}",
    eprint = "2401.11676",
    archivePrefix = "arXiv",
    primaryClass = "nucl-ex",
    doi = "10.3390/universe10040168",
    journal = "Universe",
    volume = "10",
    number = "4",
    pages = "168",
    year = "2024"
}

@article{NA64,
    author = "Banerjee, D. and others",
    collaboration = "NA64",
    title = "{Improved limits on a hypothetical X(16.7) boson and a dark photon decaying into $e^+e^-$ pairs}",
    eprint = "1912.11389",
    archivePrefix = "arXiv",
    primaryClass = "hep-ex",
    reportNumber = "CERN-EP-2019-284",
    doi = "10.1103/PhysRevD.101.071101",
    journal = "Phys. Rev. D",
    volume = "101",
    number = "7",
    pages = "071101",
    year = "2020"
}

@article{Feng1,
    author = "Feng, Jonathan L. and Fornal, Bartosz and Galon, Iftah and Gardner, Susan and Smolinsky, Jordan and Tait, Tim M. P. and Tanedo, Philip",
    title = "{Protophobic Fifth-Force Interpretation of the Observed Anomaly in $^8$Be Nuclear Transitions}",
    eprint = "1604.07411",
    archivePrefix = "arXiv",
    primaryClass = "hep-ph",
    reportNumber = "UCI-TR-2016-09",
    doi = "10.1103/PhysRevLett.117.071803",
    journal = "Phys. Rev. Lett.",
    volume = "117",
    number = "7",
    pages = "071803",
    year = "2016"
}

@article{Feng2,
    author = "Feng, Jonathan L. and Fornal, Bartosz and Galon, Iftah and Gardner, Susan and Smolinsky, Jordan and Tait, Tim M. P. and Tanedo, Philip",
    title = "{Particle physics models for the 17 MeV anomaly in beryllium nuclear decays}",
    eprint = "1608.03591",
    archivePrefix = "arXiv",
    primaryClass = "hep-ph",
    reportNumber = "UCI-TR-2016-12",
    doi = "10.1103/PhysRevD.95.035017",
    journal = "Phys. Rev. D",
    volume = "95",
    number = "3",
    pages = "035017",
    year = "2017"
}

@article{Ell,
    author = "Ellwanger, Ulrich and Moretti, Stefano",
    title = "{Possible Explanation of the Electron Positron Anomaly at 17 MeV in $^8Be$ Transitions Through a Light Pseudoscalar}",
    eprint = "1609.01669",
    archivePrefix = "arXiv",
    primaryClass = "hep-ph",
    reportNumber = "LPT-ORSAY-16-54",
    doi = "10.1007/JHEP11(2016)039",
    journal = "JHEP",
    volume = "11",
    pages = "039",
    year = "2016"
}

@article{Alves,
    author = "Alves, Daniele S. M. and others",
    title = "{Shedding light on X17: community report}",
    doi = "10.1140/epjc/s10052-023-11271-x",
    journal = "Eur. Phys. J. C",
    volume = "83",
    number = "3",
    pages = "230",
    year = "2023"
}

@article{CAPOLUPO,
    author = "Capolupo, Antonio and Quaranta, Aniello and Serao, Raoul",
    title = "{The impact of the X17 boson on particle physics anomalies: Muon anomalous magnetic moment, Lamb shift and W mass}",
    eprint = "2410.01430",
    archivePrefix = "arXiv",
    primaryClass = "hep-ph",
    doi = "10.1016/j.dark.2024.101748",
    journal = "Phys. Dark Univ.",
    volume = "47",
    pages = "101748",
    year = "2025"
}

@article{anch,
    author = "Anchordoqui, Luis A. and Antoniadis, Ignatios and Huang, Xing and Lust, Dieter and Taylor, Tomasz R.",
    title = "{Leptophilic U(1) massive vector bosons from large extra dimensions}",
    eprint = "2105.02630",
    archivePrefix = "arXiv",
    primaryClass = "hep-ph",
    reportNumber = "MPP-2021-68; LMU-ASC 12/21",
    doi = "10.1016/j.physletb.2021.136585",
    journal = "Phys. Lett. B",
    volume = "820",
    pages = "136585",
    year = "2021"
}

@article{W1,
    collaboration = "CMS",
    title = "{Measurement of the W boson decay branching fraction ratio B(W{\textrightarrow}cq)/B(W{\textrightarrow}qq{\textasciimacron}') in proton-proton collisions at s=13TeV}",
    doi = "10.1016/j.physletb.2025.139754",
    journal = "Phys. Lett. B",
    volume = "868",
    pages = "139754",
    year = "2025"
}

@article{W03,
    author = "Davoudiasl, Hooman and Enomoto, Kazuki and Lee, Hye-Sung and Lee, Jiheon and Marciano, William J.",
    title = "{Searching for new physics effects in future W mass and sin2{\ensuremath{\theta}}W(Q2) determinations}",
    eprint = "2309.04060",
    archivePrefix = "arXiv",
    primaryClass = "hep-ph",
    doi = "10.1103/PhysRevD.108.115018",
    journal = "Phys. Rev. D",
    volume = "108",
    number = "11",
    pages = "115018",
    year = "2023"
}

@article{W2,
    author = "Burgess, C. P. and Godfrey, Stephen and Konig, Heinz and London, David and Maksymyk, Ivan",
    title = "{Model independent global constraints on new physics}",
    eprint = "hep-ph/9312291",
    archivePrefix = "arXiv",
    reportNumber = "MCGILL-93-12, NEIPH-93-008, OCIP-C-93-6, UQAM-PHE-93-08, UDEM-LPN-TH-93-155",
    doi = "10.1103/PhysRevD.49.6115",
    journal = "Phys. Rev. D",
    volume = "49",
    pages = "6115--6147",
    year = "1994"
}

@article{agg1,
    author = "Capolupo, Antonio and Luongo, Orlando and Quaranta, Aniello",
    title = "{Impact of flavor condensate dark matter on accretion disk luminosity in spherical spacetimes}",
    eprint = "2507.03758",
    archivePrefix = "arXiv",
    primaryClass = "gr-qc",
    doi = "10.1140/epjc/s10052-026-15371-2",
    journal = "Eur. Phys. J. C",
    volume = "86",
    number = "2",
    pages = "154",
    year = "2026"
}

@article{agg2,
    author = "Capolupo, Antonio and Pisacane, Gabriele and Quaranta, Aniello and Serao, Raoul",
    title = "{Single arm interferometry to probe the scalar field dark matter}",
    eprint = "2505.13574",
    archivePrefix = "arXiv",
    primaryClass = "hep-ph",
    doi = "10.1103/ypm5-jh2j",
    journal = "Phys. Rev. D",
    volume = "113",
    number = "5",
    pages = "055027",
    year = "2026"
}

@article{agg3,
    author = "Capolupo, Antonio and Lambiase, Gaetano and Quaranta, Aniello",
    title = "{Fermion mixing in curved spacetime}",
    eprint = "2305.07533",
    archivePrefix = "arXiv",
    primaryClass = "hep-th",
    doi = "10.1088/1742-6596/2533/1/012050",
    journal = "J. Phys. Conf. Ser.",
    volume = "2533",
    number = "1",
    pages = "012050",
    year = "2023"
}

@article{agg4,
    author = "Capolupo, Antonio and Lambiase, Gaetano and Quaranta, Aniello",
    title = "{Neutrinos in curved spacetime: Particle mixing and flavor oscillations}",
    eprint = "2003.00516",
    archivePrefix = "arXiv",
    primaryClass = "hep-th",
    doi = "10.1103/PhysRevD.101.095022",
    journal = "Phys. Rev. D",
    volume = "101",
    number = "9",
    pages = "095022",
    year = "2020"
}

@article{agg5,
    author = "Capolupo, A. and De Martino, I. and Lambiase, G. and Stabile, An.",
    title = "{Axion{\textendash}photon mixing in quantum field theory and vacuum energy}",
    eprint = "1901.10473",
    archivePrefix = "arXiv",
    primaryClass = "hep-ph",
    doi = "10.1016/j.physletb.2019.01.056",
    journal = "Phys. Lett. B",
    volume = "790",
    pages = "427--435",
    year = "2019"
}

@article{agg6,
    author = "Feng, Jonathan L. and Tait, Tim M. P. and Verhaaren, Christopher B.",
    title = "{Dynamical Evidence For a Fifth Force Explanation of the ATOMKI Nuclear Anomalies}",
    eprint = "2006.01151",
    archivePrefix = "arXiv",
    primaryClass = "hep-ph",
    reportNumber = "UCI-TR-2020-01",
    doi = "10.1103/PhysRevD.102.036016",
    journal = "Phys. Rev. D",
    volume = "102",
    number = "3",
    pages = "036016",
    year = "2020"
}

@article{agg7,
    author = "Feng, Jonathan L. and Fornal, Bartosz and Galon, Iftah and Gardner, Susan and Smolinsky, Jordan and Tait, Tim M. P. and Tanedo, Philip",
    title = "{Particle physics models for the 17 MeV anomaly in beryllium nuclear decays}",
    eprint = "1608.03591",
    archivePrefix = "arXiv",
    primaryClass = "hep-ph",
    reportNumber = "UCI-TR-2016-12",
    doi = "10.1103/PhysRevD.95.035017",
    journal = "Phys. Rev. D",
    volume = "95",
    number = "3",
    pages = "035017",
    year = "2017"
}

@article{agg8,
    author = "Feng, Jonathan L. and Fornal, Bartosz and Galon, Iftah and Gardner, Susan and Smolinsky, Jordan and Tait, Tim M. P. and Tanedo, Philip",
    title = "{Protophobic Fifth-Force Interpretation of the Observed Anomaly in $^8$Be Nuclear Transitions}",
    eprint = "1604.07411",
    archivePrefix = "arXiv",
    primaryClass = "hep-ph",
    reportNumber = "UCI-TR-2016-09",
    doi = "10.1103/PhysRevLett.117.071803",
    journal = "Phys. Rev. Lett.",
    volume = "117",
    number = "7",
    pages = "071803",
    year = "2016"
}
\end{document}